\documentstyle[aps,pra,amssymb,graphics]{revtex}

\begin{document}

%\draft
%\widetext
\narrowtext
\twocolumn
\wideabs{
\title{Transfer of quantum states using finite resources}

\author{Dietmar G. Fischer, Holger Mack and
Matthias Freyberger} 
\address{Abteilung f\"ur Quantenphysik, Universit\"at Ulm, 
D-89069 Ulm, Germany} 

%\thanks[didi]{E-mail: fis@physik.uni-ulm.de}

\maketitle
\begin{abstract}
We discuss the problem of transfering a qubit from Alice to Bob using a noisy
quantum channel and only finite resources. As the basic protocol for the
transfer we apply quantum teleportation. It turns out that for a certain quality
of the channel direct teleportation combined with qubit purification is superior
to entanglement purification of the channel. If, however, the quality of the
channel is rather low one should simply apply an estimation-preparation scheme.
\end{abstract}
}

\section{Introduction}

Ideal quantum information protocols \cite{zeilinger} do usually not consider the
resources needed for performing a certain task. However, in quantum information
processing the question of resources automatically comes into play. The reason
for this is that the measurement process is an important part of all
considerations and each quantum measurement changes the state of the system.
Hence a quantum measurement sequence is fundamentally different from a classical
one used in classical computing (e.g. readout of a quantum register versus readout
of a classical register): Collecting information about a quantum system
\cite{helstrom} usually requires the preparation of an ensemble of those
systems.
The accuracy needed for this information typically determines the size of the
ensemble, that is, the resources for this task.\par
 Typical examples for this
situation have been discussed recently: Quantum state estimation
\cite{helstrom,massar,bruss,tarrach,fischer2}
asks for the optimal exploitation of $N$ identically prepared quantum systems in
order to find their state. This problem is closely related to optimal quantum
cloning \cite{bruss,cloning} which considers the map of a finite resource of $N$
qubits onto $M$ clones. Furthermore, the question of finite resources and
accuracy poses itself when one investigates optimal frequency standards
\cite{frequency} and optimal quantum clocks \cite{clocks}.\par
In the present paper we shall discuss the transfer of quantum states from Alice
to Bob under the assumption that they have only finite resources at hand. This
problem becomes very important when one considers concepts of distributed
quantum computing \cite{distributed} where quantum computations are performed by
spatially separated quantum processors that communicate with each other via
quantum channels. The
ideal solution to the state-transfer problem is known: quantum teleportation \cite{teleport}.
If Alice and Bob share a perfect Bell state, they can apply this protocol and
Bob will perfectly receive Alice's qubit. As a resource they just need the Bell
state which sets up the quantum channel between the two parties.\par
This ideal situation changes drastically when the quantum channel becomes noisy,
that is, when Alice and Bob only share a mixed entangled state. Then it turns out
that they have different possibilities to transfer a qubit. In the following we
shall investigate three versions. First, they can start from a resource of $N$
non-maximally entangled states and purify \cite{deutsch,bennett} them until they find a 
``relatively pure'' state for teleportation. Second, they can transfer the
qubits using the teleportation protocol with the $N$ unpurified entangled
states. The question then is whether Bob can extract the original qubit from the
$N$ mixed qubits he gets. For this purpose he can use a qubit purification
protocol \cite{cirac}. Third, Alice and Bob can perform an obvious protocol.
Alice simply estimates \cite{massar,bruss,tarrach,fischer2} the state of the qubit from the $N$
systems she possesses. Then she classically 
sends this information to Bob who can use it to
prepare a corresponding qubit.
We shall compare and contrast these methods by calculating the average
fidelities in each case.\par
The paper is organized as follows: In Sec. II we present the physical scenario
that we want to investigate and introduce the corresponding notation. Sec. III
contains the description of the three different methods for the quantum-state
transfer. These methods are then compared with each other in Sec. IV. We
conclude with Sec. V.

%%%%%%%%%%%%%%%%%%%%%%%%%%%%%%%%%%%%%%%%%%%%%%%%%%%%%%%%%%%%%%%%%%%%%%%%%%     
\section{Transfer of quantum states}

We shall investigate the following scenario: Imagine
that Alice wants to send an unknown 
quantum state $|\psi\rangle$, which will be a qubit
state in our case, to Bob. The state can be parameterized in terms of the
basis states $|0\rangle$ and $|1\rangle$ as
\begin{equation}
|\psi\rangle=\cos\frac{\theta}{2}|0\rangle+\sin\frac{\theta}{2}e^{i\phi}
|1\rangle
\end{equation}
with Bloch angles $\theta\in[0,\pi]$ and $\phi\in[0,2\pi)$. At this point the
question arises which type of quantum channel the two parties should use. In the
present paper we shall consider quantum teleportation as a possible candidate.
We assume that
Alice and Bob possess all the necessary
ingredients for doing quantum state teleportation\cite{teleport}. 
If these ingredients were perfect we could have an ideal transmission
scheme: Alice and Bob set up the quantum channel using a single copy of a
maximally entangled state, i.e., they use one of the four Bell states. Then
Alice sends her copy of $|\psi\rangle$ to Bob according to the well-known
protocol of quantum state teleportation \cite{teleport}.\par
In the present paper, 
however, we assume that the channel is noisy and hence the corresponding
entangled pair of qubits (ebit) is not given by one of the Bell states but by a
density operator.
Reasons for this non-ideal entanglement could be decoherence, an imperfect
ebit source or imperfect transfer channels from the ebit source to Alice and
Bob.\par
Consequently, the state of
Bob's qubit will differ from $|\psi\rangle$ due to the imperfect ebit and has
to be described by a mixed state. However, as Bob 
wants to use his qubits for
further experiments his aim will be to possess at least one qubit in a state as
close as possible to the the initial state $|\psi\rangle$. In order to achieve
this goal Alice can send not only one copy of $|\psi\rangle$ but has $N$ copies
available which she can send to Bob. This, of course, also requires that Alice
and Bob can set up the channel with $N$ ebits.
Furthermore, Bob is allowed to perform any operations on his qubits. 
His final
state will then be called $\hat{\rho}_{B}$.\par
The quality of Bob's state will be measured in terms of the fidelity 
$\langle\psi|\hat{\rho}_{B}|\psi\rangle$ 
which quantifies the overlap between the initial and the final state.
Our question therefore is how we can improve the fidelity using the given
finite resources, i.e., using the $N$ ebits for the channel and the $N$ copies
of $|\psi\rangle$ which Alice can send.\par
Before we consider the different strategies with which Bob can improve the
state $\hat{\rho}_{B}$, i.e., improve the fidelity, 
let us first look at Bob's state 
resulting from a teleportation with a mixed ebit state. 
We assume that the ebit setting up the channel is in a Werner state
\cite{werner}
\begin{eqnarray}
\hat{\rho}_e(\lambda)&=&\lambda |\phi^+\rangle\langle\phi^+|\nonumber \\
&+&\frac{1-\lambda}{3}
\left(|\psi^+\rangle\langle\psi^+|+|\phi^-\rangle\langle\phi^-|
+|\psi^-\rangle\langle\psi^-|\right)
\end{eqnarray}
with the parameter $\lambda\in[1/4,1]$ and the four Bell states
$|\phi^{\pm}\rangle$ and $|\psi^{\pm}\rangle$ \cite{zeilinger}. 
The parameter $\lambda$ defines the quality of the channel. For $\lambda=1$ we
recover the ideal teleportation again whereas for $\lambda=1/4$ the Werner state
is completely mixed and consequently no teleportation is possible. 
Note that the assumption of a Werner state does not 
restrict the generality of our results because any ensemble of quantum states
can be converted into a Werner state by applying local random rotations onto the
two qubits of the ebit. \par
As a result of the teleportation Bob will get the state 
\begin{equation}
\hat{\rho}_B(\lambda)=\frac{1+2\lambda}{3}|\psi\rangle\langle\psi|+
\frac{2}{3}(1-\lambda)|\bar{\psi}\rangle\langle\bar{\psi}|,
\end{equation}
where the state 
\begin{equation}
|\bar{\psi}\rangle=\sin\frac{\theta}{2}|0\rangle-\cos\frac{\theta}{2}
e^{i\phi}|1\rangle
\end{equation} 
is orthogonal to $|\psi\rangle$. 
That is, we find a classical mixture of $|\psi\rangle$ and $|\bar{\psi}\rangle$
in which the desired $|\psi\rangle$ component prevails if the ebit parameter
fulfills the condition 
\begin{equation}
\lambda>\lambda_{crit}=\frac{1}{4}.
\end{equation}
From this representation we can easily determine 
the fidelity  
\begin{equation}
F_B(\lambda)=\langle\psi|\hat{\rho}_B(\lambda)|\psi\rangle=\frac{2\lambda+1}{3}.
\end{equation}
of Bob's output state.
This fidelity is of course the fidelity that we get if we run the imperfect
teleportation apparatus only once. However, we have $N$ ebits and $N$ qubits
at our
disposal and thus can perform the teleportation at most $N$ times. Therefore, 
we can
ask the question: What is the best way to use our finite resources to achieve an
output state $\hat{\rho}_{B}$ 
as close as possible to the initial state $|\psi\rangle$? 
The goal of the following sections will be to discuss and compare several 
methods designed for this purpose. 

%%%%%%%%%%%%%%%%%%%%%%%%%%%%%%%%%%%%%%%%%%%%%%%%%%%%%%%%%%%%%%%%%%%%%%%%%%%%%%%
\section{Methods to improve the transfer fidelity}

In this section we will discuss three methods with which Alice and Bob can
improve the fidelity of the final output state $\hat{\rho}_{B}$. 
To some of these methods there exist a number of related versions. We will,
however, concentrate on only one specific method in each case.
 
%%%%%%%%%%%%%%%%%%%%%%%%%%%%%%%%%%%%%%%%%%%%%%%%%%%%%%%%%%%%%%%%%%%%%%%%%%%%%
\subsection{Entanglement purification}

The first strategy to improve the transport quality of $|\psi\rangle$ 
is to use one of the
existing entanglement purification protocols to get one highly entangled pair of
qubits out of the available $N$ mixed ebits. This final pair of qubits can then
be used to teleport $|\psi\rangle$ from Alice to Bob.
As an example of such a protocol we will
use the one proposed by Deutsch et al. \cite{deutsch}, which for Werner states
yields the same results as the original entanglement purification
protocol\cite{bennett}, but is conceptually simpler and can also be applied
to ebits in a general Bell-diagonal state. This purification
scheme works for all Werner states with
$\lambda >1/2$.
In addition, the purification protocol 
is quite simple with respect to experimental
realizability because it only requires a low number of fundamental quantum
operations to be performed.  \par
The purification scheme of Deutsch et al. \cite{deutsch} consists of 
the following
steps. First, 
Alice and Bob both take the qubits of a pair of ebits and perform a unitary
operation on each of these qubits separately. Second, they
perform a controlled-not operation on their pairs of qubits and make
sure that for each of them the control and target qubits belong to the same
ebit.
Third, 
Alice and Bob measure the target qubits in the computational basis
$|0\rangle$ and $|1\rangle$
and tell each other the results. If the results coincide they keep the
corresponding control
qubits. The control qubit of Bob and the control qubit of Alice form the
purified ebit. Otherwise, if their measurement results are not the same, they
have to discard their control qubits. This, of course, means that the
purification was not successful and hence Bob and Alice lost two ebits.\par
If the purification steps are successful
one finds that for two Werner states, Eq.(2), with
parameter $\lambda$ the resulting density operator of the purified ebit reads
\begin{eqnarray} 
\hat{\rho}&=&\frac{1}{9 p_{pass}}\left(10\lambda^2-2\lambda
+1\right)|\phi^+\rangle\langle\phi^+|\nonumber \\
&+&\frac{2}{3 p_{pass}}\left(\lambda-\lambda^2\right)
|\phi^-\rangle\langle\phi^-| \nonumber \\
&+&\frac{2}{9 p_{pass}}(1-\lambda)^2
\left(|\psi^+\rangle\langle\psi^+|+|\psi^-\rangle\langle\psi^-|
\right)
\end{eqnarray}
with the probability 
\begin{equation}
p_{pass}=\frac{1}{9}\left(8\lambda^2-4\lambda+5\right)
\end{equation}
to pass the measurement test at the end of the purification scheme.
Of course, this is no longer a Werner state but can be transformed into one by
local random rotations again \cite{bennett}. \par
As the aim of our purification scheme is the preparation of one highly entangled
pair we propose the following algorithm that is also depicted in Fig. 1.
We start with our finite set of $N$ initial ebits, all in the same Werner state
$\hat{\rho}_e(\lambda_0)$, Eq.(2), with $\lambda_0> 1/4$. We recall that
$\lambda_0=\lambda_{crit}=1/4$ was the critical value which limits the use of
teleportation as a transport channel for the state $|\psi\rangle$. It is,
however, obvious that purification of our ebits will not improve the fidelity of
Bob's output state if we start in the range $1/4<\lambda_0\le 1/2$. The reason
for this simply is that the purification protocol does not work in this range
since $\hat{\rho}_e(\lambda_0)$, Eq.(2), is then separable. We will, however, include
this interval in our calculations in order to see explicitely how the output
fidelity of Bob's state changes with growing $\lambda_0$.\par
Let us now go through the steps of our repeated purification algorithm, see
Fig. 1: 

If $i=N$ is even, we directly perform the first purification step
using all $i$ ebits. For odd $i$ we first store one ebit and then continue with
the purification using only $i-1$ ebits. The result of the $n$-th 
purification step performed on $i$ Werner states with parameter $\lambda_{n-1}$
will be a number of $j\in \{0,1,...,i/2\}$ successfully purified ebits, where
each possible $j$ occurs with the probability
\begin{equation}
p_{ij}(\lambda_{n-1})={i/2 \choose j}
p_{pass}^j(\lambda_{n-1})[1-p_{pass}(\lambda_{n-1})]^{i/2-j}.
\end{equation}
The purified ebits can now be converted into Werner states with the parameter
\begin{equation}
\lambda_n=\frac{10\lambda_{n-1}^2-2\lambda_{n-1}+1}{8\lambda_{n-1}^2
-4\lambda_{n-1}+5}
\end{equation}
after the $n$-th purification step.
If we still have at least two ebits left, i.e., $j>1$ 
we repeat our purification using
the already purified ebits. For
$j=1$ we only have one ebit left and use it to teleport our initial qubit to
Bob. The resulting fidelity of Bob's qubit in state $\hat{\rho}_B(\lambda_n)$,
Eq.(3), 
will then be
$F_B(\lambda_n)$, Eq.(6),
if the purification has been performed $n$ times. In the case of $j=0$, however,
we have to look for previously stored ebits. If we have stored ebits we use the
lastly stored ebit for the teleportation \cite{entpur}. 
The teleportation fidelity will be   
$F_B(\lambda_k)$, if we have stored the last ebit 
after the $k$-th purification step. 
The worst case occurs when we have no stored ebits at all. In this case we have
lost all our ebits and thus cannot use them for teleportation. As Bob has no
information about the initial qubit state, he can only achieve a fidelity
$F_B(1/4)=1/2$. That is, his information must be described by a completely mixed
state.\par
Hence, if we start with $N$ ebits defined by $\lambda_0$ we can now calculate,
using the algorithm above, the average fidelity
\begin{equation}
F_B^{(1)}(N,\lambda_0)=\Big\langle F_B\Big\rangle
\end{equation}
of Bob's output state. Note that the averaging $\langle ... \rangle$ here means
to average over all possible paths through the algorithm depending on the
probabilities, Eq.(9).
The resulting average fidelities $F_B^{(1)}(N,\lambda_0)$ of our purification scheme 
are shown in Fig. 2. As one would expect
the fidelities always increase with growing $\lambda_0$ and approach 1 for
$\lambda_0\rightarrow 1$.\par
 The dependency of $ F_B^{(1)}(N,\lambda_0)$ on $N$ is
more complicated. As expected the fidelity shows the behaviour 
$F_B^{(1)}(N=1,\lambda_0)>F_B^{(1)}(1<N\le 32,\lambda_0)$ in the range
$1/4<\lambda_0<1/2$.
There the
Werner state is separable and the entanglement purification method yields no
improvement as argued before. Or, in other words, in this $\lambda_0$ range one
could simply use one of the unpurified states $\hat{\rho}_e(\lambda_0)$ 
and perform the
teleportation with it. On the other hand if 
we start with a larger initial Werner parameter
$\lambda_0>1/2$ we clearly see an increase of $F_B^{(1)}$ with growing $N$.
\par
However, there is a clear difference between even and odd $N$. The fidelities
for odd $N$ are always considerably higher than in the case of adjacent 
even $N$. This is
a consequence of the fact that we never lose all the ebits for odd $N$. Thus
we see that on average we get a better quality of Bob's output state if we only
use the highest odd number of ebits for the entanglement purification. On the
other hand this means that for an even $N$ one should discard one ebit first
and perform the purification with the remaining $N-1$ ebits. For this reason we
will always use this modified method for the remaining parts of the paper, i.e.,
in the case of even $N$ we will only use $N-1$ ebits for the 
entanglement-purification method so that the effective average fidelity is 
$F_B^{(1)}(N-1,\lambda_0)$ for even
$N$. \par

%%%%%%%%%%%%%%%%%%%%%%%%%%%%%%%%%%%%%%%%%%%%%%%%%%%%%%%%%%%%%%%%%%%%%%%%%%%%%
\subsection{Qubit purification}

Instead of purifying the $N$ ebits that are used for the teleportation Alice 
could simply use all of the $N$ ebits in state $\hat{\rho}_e(\lambda_0)$ 
to teleport $N$ states $|\psi\rangle$ so
that Bob would get $N$ qubits in the state $\hat{\rho}_B(\lambda_0)$, Eq. (3). 
Bob can then apply
a qubit purification protocol \cite{cirac} 
to his $N$ states described by the product $\hat{\rho}_B^{\otimes N}\equiv
\hat{\rho}_B^1(\lambda_0)\otimes ...\otimes\hat{\rho}_B^N(\lambda_0)$. 
The qubit purification
protocol performs a projection of the qubit product state
$\hat{\rho}_B^{\otimes N}$ which yields an entangled state $\hat{\rho}_M$ made
up of $M$ qubits and a product state of Bell-$|\psi^-\rangle$ states. Thus the
effective transformation of the qubit purification procedure can be written as
\begin{equation}
P:\hat{\rho}_B^{\otimes N} \mapsto  \hat{\rho}_M \otimes \left(|\psi^-\rangle
\langle\psi^-|\right)^{\otimes (N-M)/2}.
\end{equation}
For even $N$ values one obtains even values 
of $M\in \{0,2,...,N\}$, 
whereas for odd $N$ one finds odd
$M\in \{1,3,...,N\}$. Any value of $M$ is obtained with a probability
\begin{equation}
p_M(\lambda_0)=d_M \left[c_0 c_1\right]^{(N-M)/2}
\frac{c_1^{M+1}-c_0^{M+1}}{c_1-c_0}
\end{equation}
where we used the notation
$c_1\equiv(1+2\lambda_0)/3$, $c_0\equiv 2(1-\lambda_0)/3$ and the combinatorical
prefactor 
\begin{equation}
d_M=\left\{ \begin{array}{ccc}
{N\choose \frac{N-M}{2}}-{N\choose \frac{N-M}{2}-1} & \mbox{for} & M<N \\
1 & \mbox{for} & M=N
\end{array} \right. .
\end{equation}
The density operator $\hat{\rho}_M$ can also be calculated \cite{cirac} 
and one finds
\begin{eqnarray}
\hat{\rho}_M(\lambda_0)&=&\frac{c_1-c_0}{c_1^{M+1}-c_0^{M+1}}(M+1)\nonumber\\
&\times&\int \frac{d\Omega'}{4\pi} \left(|\Psi(\theta',\phi')\rangle\langle
\Psi(\theta',\phi')|\right)^{\otimes M}
\end{eqnarray}
where the unnormalized states
\begin{equation}
|\Psi(\theta',\phi')\rangle=\sqrt{c_1}\cos\frac{\theta'}{2}|\psi\rangle+
\sqrt{c_0}\sin\frac{\theta'}{2}e^{i\phi'}|\bar{\psi}\rangle
\end{equation}
are a superposition of the original qubit state $|\psi\rangle$, Eq.(1), 
and the corresponding orthogonal state 
$|\bar{\psi}\rangle$, Eq.(4). \par
After having performed the qubit purification procedure we discard the 
$N-M$ qubits
in the $|\psi^-\rangle$ state and just keep the $M$ qubits in the entangled state 
$\hat{\rho}_M$. The final goal of our scheme is to get one output
qubit with a maximal fidelity compared to the initial state $|\psi\rangle$.
Thus we have to look at
the reduced density operator $\hat{\rho}^{red}_M(\lambda_0)$ 
of $\hat{\rho}_M(\lambda_0)$ that can be
evaluated by tracing over all qubits except of one. The average fidelity of
$\hat{\rho}_M^{red}$ then reads \cite{cirac}
\begin{eqnarray}
f_M(\lambda_0)&=&\langle\psi|\hat{\rho}^{red}_M(\lambda_0)|\psi\rangle
\nonumber \\
&=&
\left\{ \begin{array}{ccc}
\frac{1}{M}\left[\frac{(M+1)c_1^{M+1}}{c_1^{M+1}-c_0^{M+1}}
-\frac{c_1}{c_1-c_0}\right] & \mbox{for} & M>0 \\
\frac{1}{2} & \mbox{for} & M=0 
\end{array} \right. .
\end{eqnarray}
This fidelity $f_M$ 
is larger than the initial fidelity $F_B(\lambda_0)$ for $1/4<\lambda_0<1$ and
$M>0$. This means that we have improved the quality of the output qubit by the
qubit purification. In contrast to the entanglement purification scheme the qubit
purification leads to an improvement for all parameters $\lambda_0>1/4$. 
The average fidelity of our output qubit for this second method 
is then given by
\begin{equation}
F_B^{(2)}(N,\lambda_0)=\left\{\begin{array}{l@{\quad : \quad}l}
\sum_{i=0}^{N/2} p_{2i}f_{2i} & N \mbox{ even}\\
\sum_{i=0}^{(N-1)/2} p_{2i+1}f_{2i+1} & N \mbox{ odd}
\end{array}\right. ,
\end{equation}
with the probabilities $p_M=p_M(\lambda_0)$, Eq.(13), and the single qubit
fidelities $f_M=f_M(\lambda_0)$, Eq.(17).
The resulting average fidelities are plotted in Fig. 3. In contrast
to the entanglement purification scheme the best performance is always obtained
by using all available qubits here.

%%%%%%%%%%%%%%%%%%%%%%%%%%%%%%%%%%%%%%%%%%%%%%%%%%%%%%%%%%%%%%%%%%%%%%%%%%%%%
\subsection{State estimation and preparation}

Alice and Bob have a third possibility to transfer a qubit state. This
possibility consists of two very basic ingredients and avoids any quantum
teleportation at all. Alice simply has to perform measurements on her $N$ quantum
systems and to estimate the quantum state from her measurement results. Then she
tells Bob the parameters of her estimated quantum state via a classical
communication channel. Bob can then prepare the qubits in the desired quantum
state on his side. For qubits the state $|\psi\rangle$, Eq.(1), can be described
by the two Bloch-parameters $(\theta,\phi)$ which Alice has to tell Bob. 
This straightforward scheme avoids the use of a noisy quantum
channel. \par 
However, Alice cannot accurately estimate the quantum state 
from a finite ensemble of $N$ qubits. The optimal state estimation limit has
been found\cite{massar} and yields the optimal estimation fidelity
\begin{equation}
F_{B}^{(3)}(N)=\frac{N+1}{N+2}.
\end{equation}
This is, of course, also the fidelity with which Bob can then prepare the
corresponding output state. Note that this fidelity does not depend on
$\lambda_0$ since no quantum channel is involved in this method. 
In addition this optimal estimation scheme requires a
simultaneous joint measurement to be performed on all $N$ qubits \cite{fischer}.

%%%%%%%%%%%%%%%%%%%%%%%%%%%%%%%%%%%%%%%%%%%%%%%%%%%%%%%%%%%%%%%%%%%%%%%%%%%  
\section{Comparison of the methods}

For the transfer of a qubit state Alice and Bob will
of course try to use the most efficient of the three methods described above.
This efficiency will be measured in terms of the average fidelity of Bob's final
output state  with respect to the initial input state
$|\psi\rangle$. 
The fidelity will in general 
depend on the quality of the quantum channel which will be
characterized by the parameter $\lambda_0$ of the ebit state
$\hat{\rho}_e$, Eq.(2), and on the number $N$ of available ebits. \par
Before we start with the more general comparison let us first look at a typical
example of the behaviour of the three proposed methods, namely for $N=9$. The
average fidelities of the output state are plotted versus $\lambda_0$ in Fig. 4.
 As
the estimation-preparation method does not depend on the quantum channel its
fidelity stays constant over the whole range of $\lambda_0$ values. The 
qubit-purification scheme increases the average fidelity for any value
$1/4<\lambda_0<1$ and offers a better result than the entanglement-purification
scheme for $N=9$ in the whole range. In addition to that we find crossing points of the
two purification schemes and the estimation-preparation method and denote them
by
$\lambda_0^{(1)}$
and $\lambda_0^{(2)}$, respectively. This means that for quantum channels with 
$\lambda_0<\lambda_0^{(2)}$
the estimation-preparation method yields better results than the 
qubit-purification method. Moreover, the qubit-purification method is always
superior to the entanglement-purification method in the case of $N=9$. 
\par
This rises the question how the crossing points change with $N$. The answer to
this question is shown in Fig. 5. The crossing points $\lambda_0^{(1)}$ and 
$\lambda_0^{(2)}$
lie at the same position for $N=1$ and also for $N=2$ \cite{why}. 
For larger $N$ we find that $\lambda_0^{(2)}$ is
always smaller than $\lambda_0^{(1)}$, again indicating the better results of the
qubit-purification scheme. Moreover, the value of $\lambda_0^{(1)}$ increases with
growing $N$, whereas $\lambda_0^{(2)}$ asymptotically converges towards the
value $5/8$ \cite{asymptotic}.
Thus we can conclude that Alice and Bob should use the 
qubit-purification method for their quantum state transfer whenever their quantum
channel yields ebits with a Werner-state parameter larger than 
$\lambda_0^{(2)}$. If they, however, only get ebits with
$\lambda_0<\lambda_0^{(2)}$ then Alice should estimate her quantum state and only send her
result to Bob via a classical channel. 
As the entanglement-purification scheme never works better than
the qubit-purification methods, its application should be avoided in any
case.\par
Due to the averaging process
that is necessary to calculate $F_B^{(1)}$, Eq.(11), and 
$F_B^{(2)}$, Eq.(18), it is hard to see this superiority of the
qubit-purification scheme from the analytic expressions. However, the higher
quality
of the presented 
qubit-purification scheme can be seen qualitatively from basic properties
of the purification schemes. The number of possible purification results
that correspond to a failure of the purification process is small for the
qubit-purification scheme \cite{cirac}. 
In this scheme only the outcome $M=0$ that occurs
with probability $p_0$, Eq.(13), leads to a failure of the purification. For the
entanglement-purification scheme, on the other hand, several sequences of
purification steps result either in a loss of all ebits or no improvement due to
the purification. 
This is even true for our scheme in which we take into account the possibility to
store ebits during the purification process. Hence the presented scheme will be
superior to standard entanglement-purification schemes which are known to have
rather low ouput yields \cite{bennett,briegel}. 
It seems to be impossible to construct an
efficient entanglement purification algorithm based on finite resources due to
the low efficiency of every single purification step.
In
contrast to this the qubit-purification scheme can lead to much larger
improvements in the output fidelities, Eq.(17), because it is only a one-step
process.

\section{Conclusion}
In this paper we have studied the problem of how to transfer a qubit state
efficiently from a sender to a receiver when only finite resources are
available. In this context we investigated three
different methods to achieve this transfer. It turned out that one can never
achieve better results by an entanglement-purification scheme than by
qubit purification. Furthermore, there exists a threshold for the quality of the
quantum channel through which the qubits are sent. If the quality of the channel
is below this threshold then Alice should estimate her quantum states and tell
Bob the results classically 
so that he can prepare the quantum state himself. If the channel
quality is above the threshold then Alice should send her qubits to Bob without
purifying the entanglement of the channel. Rather Bob
should apply the qubit-purification method to get his final output state.\par
This state transfer problem can achieve practical importance in the context 
of distributed quantum computing \cite{distributed} where quantum states have to
be exchanged between spatially separated quantum processors. As there
is a cost (computation time and number of resources) associated with each
computational step, it is extremely important to use improved protocols for the
quantum state transfer in this case.

%%%%%%%%%%%%%%%%%%%%%%%%%%%%%%%%%%%%%%%%%%%%%%%%%%%%%%%%%%%%%%%%%%%%%%%%%%    
\acknowledgments
We acknowledge support by the DFG programme
``Quanten-Informationsverarbeitung'',
by the European Science Foundation QIT programme and by the programmes
``QUBITS'' and ``QUEST'' of the European Commission.  

%%%%%%%%%%%%%%%%%%%%%%%%%%%%%%%%%%%%%%%%%%%%%%%%%%%%%%%%%%%%%%%%%%%%%%%%%%

\newpage

%%%%%%%%%%%%%%%%%%%%%%%%%%%%%%%%%%%%%%%%%%%%%%%%%%%%%%%%%%%%%%%%%%%%%%%%

\begin{figure}
\vspace{3cm}
\resizebox{8cm}{!}{\includegraphics[95,-50][550,800]{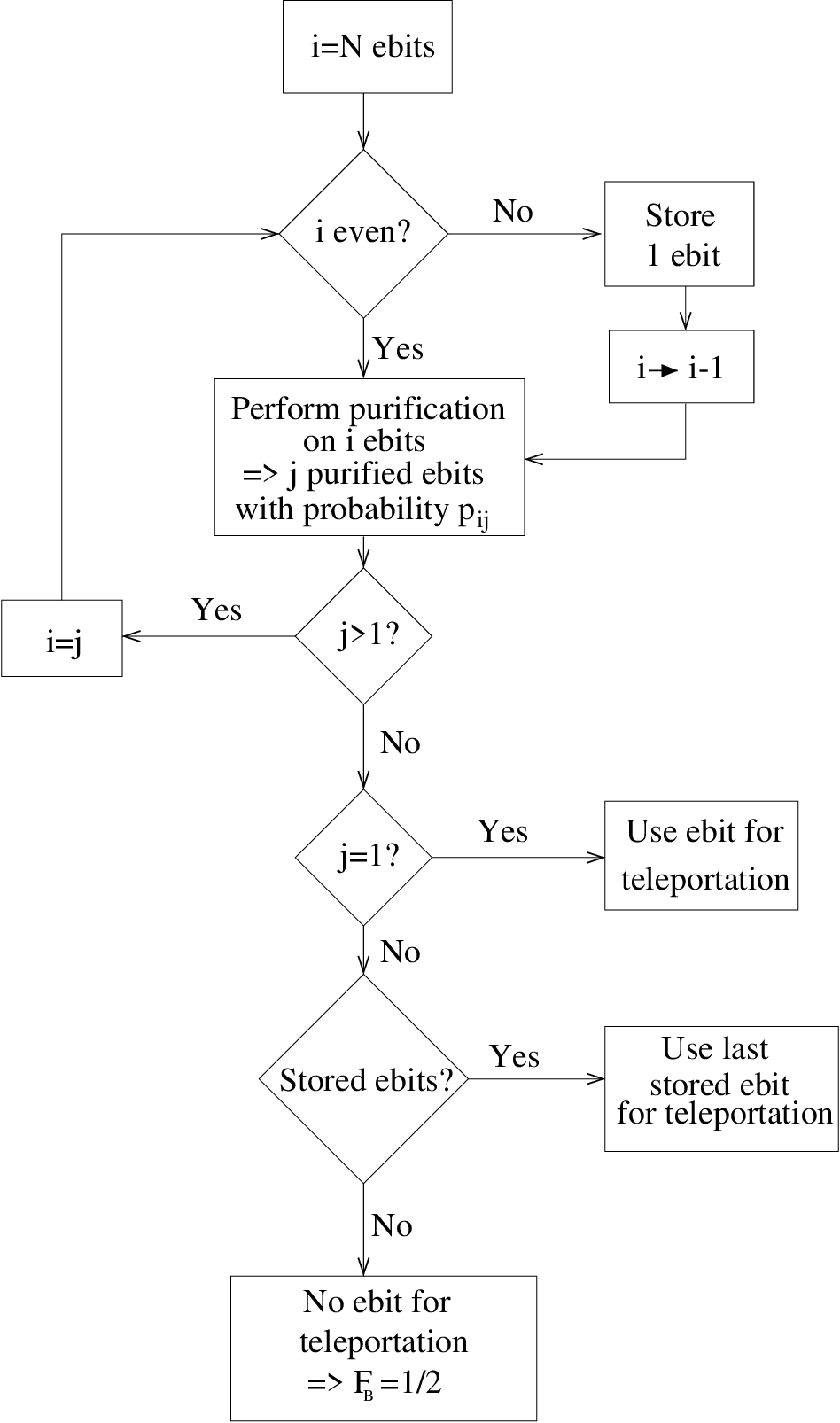}}
\caption{Flow chart describing the repeated 
entanglement-purification algorithm. The aim of
the algorithm is to generate one highly entangled ebit from an initial supply of
$N$ ebits. If we get one purified ebit in the last purification step, i.e.,
$j=1$ then we use this ebit for the teleportation. If no purified ebit is left
over ($j=0$) we look for previously stored ebits. The ebit that has been stored
lastly is then used for the teleportation. Only if no ebit has been stored
during the purification procedure we have no ebit available for the
teleportation. Thus no qubit state can be transfered to Bob and the fidelity of
his output qubit will be 1/2.}
\end{figure} 

\begin{figure}
\resizebox{8cm}{!}{\includegraphics[95,100][550,650]{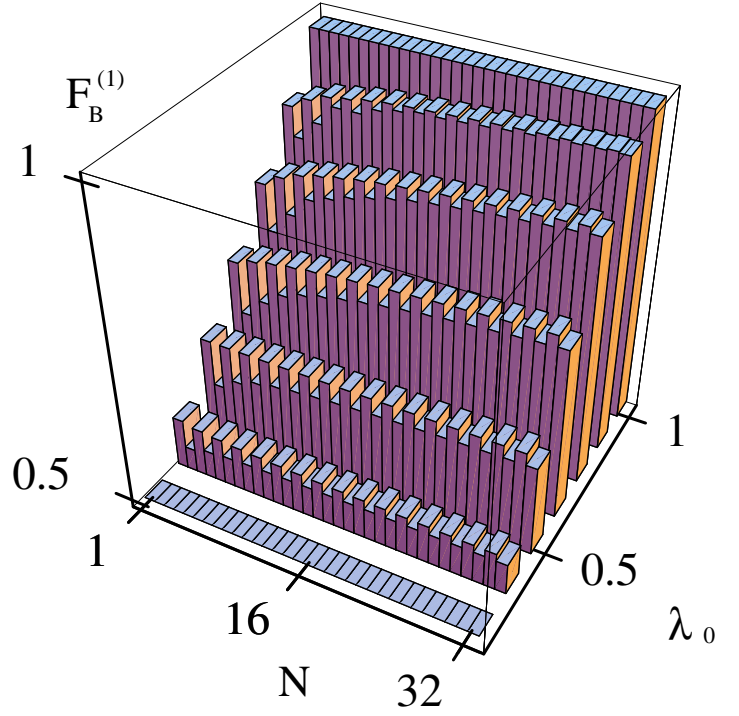}}
\caption{Plot of Bob's qubit fidelity $F^{(1)}_B$ resulting from an application
of the entanglement-purification method. The dependency of $F^{(1)}_B$ on the
number $N$ of available ebits and the quantum channel quality, represented by
$\lambda_0$, is shown. For all $\lambda_0\in(0.25,1)$
the fidelities for odd $N$ are higher than for the adjacent even values. In
addition, the fidelities decrease with growing 
$N$ for $\lambda_0<0.5$ and increase with $N$ for $\lambda_0>0.5$. As expected
the output fidelity always improves if the parameter $\lambda_0$ is increased.}
\end{figure} 

\begin{figure}
\resizebox{8cm}{!}{\includegraphics[95,100][550,650]{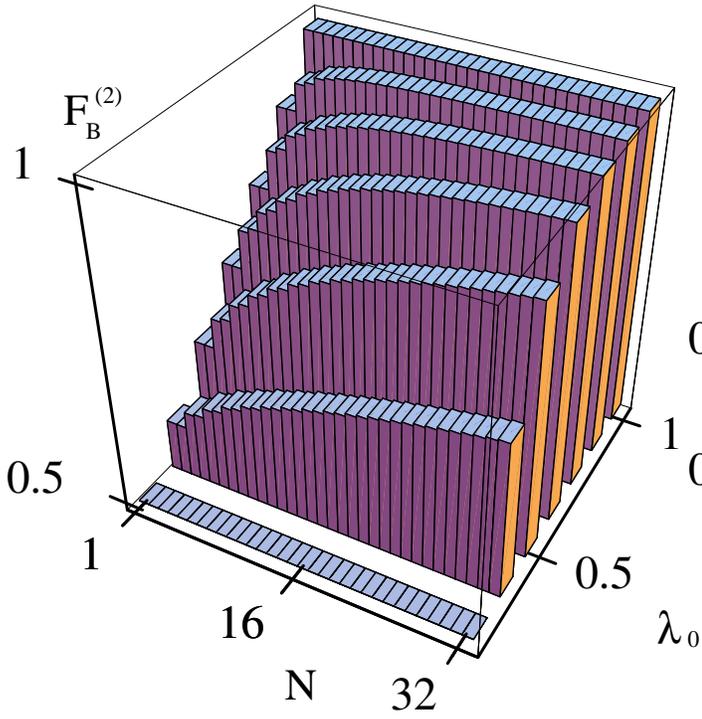}}
\caption{Output fidelity $F^{(2)}_B$ of Bob's qubit after applying the 
qubit-purification scheme. The fidelity is 
plotted versus the number $N$ of available qubits and
the parameter $\lambda_0$. In contrast to the
entanglement-purification method, cf. Fig. 2, $F^{(2)}_B$ always increases with
growing $N$ and $\lambda_0$ for $N>2$ and $1/4<\lambda_0<1$. Note the remarkably
large increase of the fidelity that already occurs for small $N$.}
\end{figure} 
\newpage
\begin{figure}
\resizebox{8cm}{!}{\includegraphics[95,100][550,650]{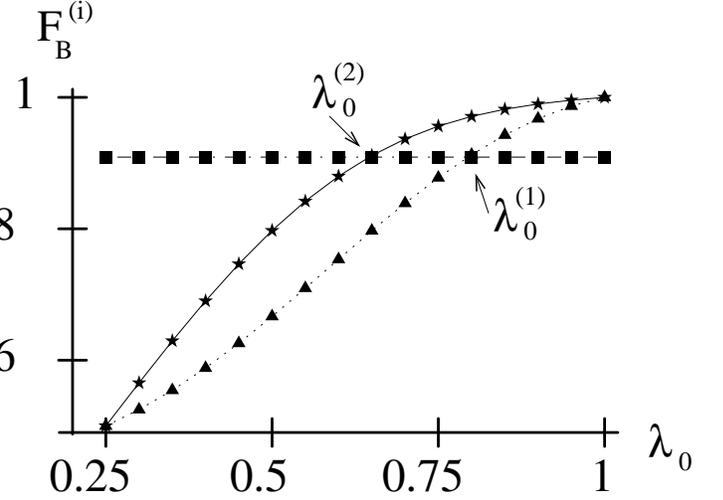}}
\caption{Comparison of the three state transfer schemes for $N=9$. Over
the whole range of possible $\lambda_0$ values the fidelities $F^{(2)}_B$
($\star$) of the qubit-purification method are always bigger than the fidelities
$F^{(1)}_B$ ($\blacktriangle$) of the entanglement-purification method. As the
fidelity $F^{(3)}_B$ ($\blacksquare$) of the estimation-preparation method does not
depend on the channel quality it is represented by a constant line that crosses
the curves for methods based on purification: We denote the crossing point of 
$F^{(1)}_B$ with $F^{(3)}_B$ by $\lambda_0^{(1)}$ and the one of $F^{(2)}_B$ 
with $F^{(3)}_B$ by $\lambda_0^{(2)}$. Obviously, for
$\lambda_0$ values smaller than the crossing-point value $\lambda_0^{(2)}$ the
estimation-preparation method should be used for transfering the state. For
larger $\lambda_0$ values the qubit-purification method should be used since it
always results in higher fidelities than the entanglement-purification method.
We emphasize that this general behavior, shown here for $N=9$, is generic, see
Fig. 5.}
\end{figure} 

\begin{figure}
\resizebox{8cm}{!}{\includegraphics[95,100][550,650]{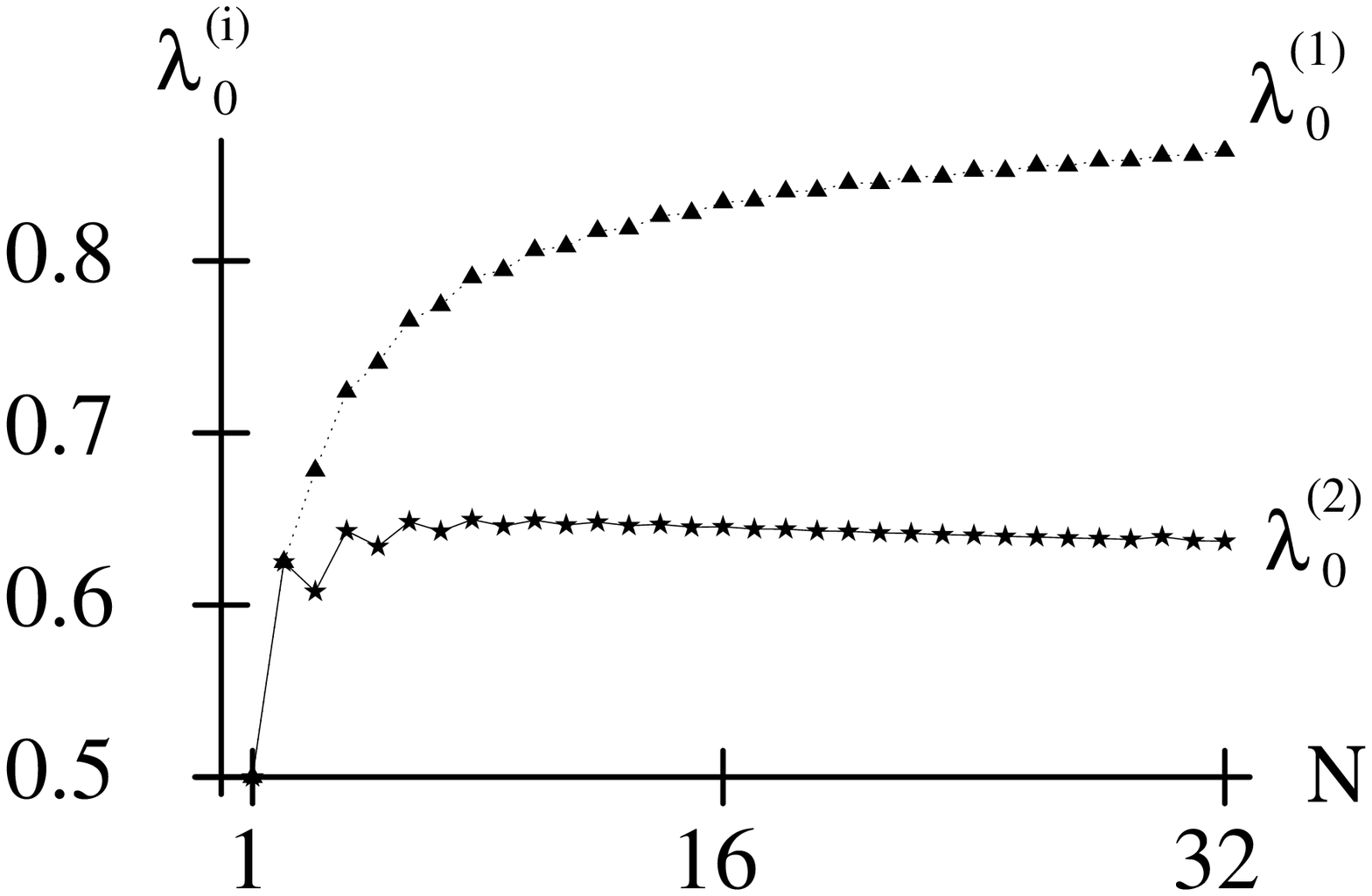}}
\caption{Plot of the crossing point values $\lambda_0^{(i)}$ versus number $N$
of available ebits. The crossing point values $\lambda_0^{(1)}$
($\blacktriangle$) of the entanglement-purification scheme increases
monotonically with $N$. For $N>2$ the values are always bigger than the
corresponding values $\lambda_0^{(2)}$ ($\star$) 
for the qubit-purification scheme.
Moreover, $\lambda_0^{(2)}$ stays almost constant for $N>6$.}
\end{figure} 

\begin{thebibliography}{99}
\bibitem{entpur} We have 
also considered the possibility to use some of the
previously stored ebits and purify further. It turns out, however, that the
average fidelity $F_B^{(1)}$ of the output qubit cannot be increased by further purification
steps of already stored ebits.
\bibitem{fischer} This disadvantage can be overcome by restricting oneself to simple spin
measurements in connection with an adaptive algorithm. The
adaptive algorithm is used to select the direction in which the spin
is measured from measurement to measurement. With such schemes 
fidelities very close to the
optimal one, Eq.(19), can be achieved. For details see Ref.
\cite{fischer2}.
\bibitem{why} We recall that due to the chosen strategy (see Sec. III.A) we
effectively have the average fidelity $F_B^{(1)}(N=1,\lambda_0)$ for the case of
$N=2$ initial ebits. Also for the qubit-purification method there is no
difference on average between $N=1$ and $N=2$ as can be seen from Fig. 3.
\bibitem{asymptotic} Taking the asymptotic expressions 
\cite{cirac} for the average fidelity of qubit purification 
one can easily calculate the asymptotic value of $\lambda_0^{(2)}$.
As result we find $\lambda_0^{(2)}\to 5/8=0.625$  for $N\to\infty$.
\end{thebibliography}
\end{document}